\documentclass[a4paper,12pt]{amsart}
\usepackage{amssymb,amsmath,mathrsfs}
\usepackage[usenames,dvipsnames]{xcolor}
\usepackage{hyperref}
\usepackage{tensor}
\usepackage{graphicx}
\usepackage{enumerate}
\usepackage{amsmath}
\usepackage{float} % positioning figures and tables
\usepackage{amsfonts}
\usepackage{amssymb}
\usepackage{enumitem}
\usepackage{physics}
\usepackage{xcolor} % for colour text
\usepackage{graphicx}

\usepackage[left=1.2in,top=1in,right=1.2in,bottom=1in,headheight=0.8in,foot=0.5in]{geometry}
\usepackage[nodisplayskipstretch]{setspace} 
\usepackage{mdframed}

\usepackage[greek.polutoniko,english]{babel}

\hypersetup{
	colorlinks=true,         
	linkcolor=MidnightBlue,          
	citecolor=MidnightBlue,
	urlcolor=MidnightBlue            
}
\let\oldmarginpar\marginpar
\renewcommand\marginpar[1]{\oldmarginpar{\color{red}\raggedright\scriptsize #1}}
\let\oldmarginpar\marginpar
\renewcommand\marginpar[1]{\oldmarginpar{\color{red}\raggedright\scriptsize #1}}

\usepackage{natbib}
\setcitestyle{aysep={}} % author date

\title[\sc Why Did the Dark Matter Hypothesis Supersede Modified Gravity in the 1980s?]{\sc Why Did the Dark Matter Hypothesis Supersede Modified Gravity in the 1980s?}
\usepackage{amsaddr}

\author{Antonis Antoniou}
\address{\vspace{-0.8pc}National and Kapodistrian University of Athens}
\email{antantoniou@phs.uoa.gr}
\thanks{This article appears in \textit{Studies in History and Philosophy of Science}, 112, pp.141-152.\\For citing purposes please refer to the published version at:\\ \url{https://doi.org/10.1016/j.shpsa.2025.06.006}}

\makeatletter
\let\uppercasenonmath\@gobble

\begin{document}
	\setstretch{1}
	\maketitle
	
	\begin{abstract}
		In the 1960s and 1970s a series of observations and theoretical developments highlighted the presence of several anomalies which could, in principle, be explained by postulating one of the following two working hypotheses: (i) the existence of dark matter, or (ii) the modification of standard gravitational dynamics in low accelerations. In the years that followed, the dark matter hypothesis as an explanation for dark matter phenomenology attracted far more attention compared to the hypothesis of modified gravity, and the latter is largely regarded today as a non-viable alternative. The present article takes an integrated history and philosophy of science approach in order to identify the reasons why the scientific community mainly pursued the dark matter hypothesis in the years that followed, as opposed to modified gravity. A plausible answer is given in terms of three epistemic criteria for the pursuitworthiness of a hypothesis: (a) its problem-solving potential, (b) its compatibility with established theories and the feasibility of incorporation, and (c) its independent testability. A further comparison between the problem of dark matter and the problem of dark energy is also presented, explaining why in the latter case the situation is different, and modified gravity is still considered a viable possibility.
	\end{abstract}

	\tableofcontents
	\setstretch{1.4}
	
	\newpage
	\section{Introduction}\label{Intro}

In 1983, Mordehai Milgrom published a series of three papers (\citeyear{milgrom1983a,milgrom1983b,milgrom1983c}) in which he developed a theory of modified gravity to accommodate a series of observational anomalies in the dynamics of galaxies that mainly appeared in the 1960s and 1970s. Milgrom's theory, which would later become known as MOND (MOdified Newtonian Dynamics), appeared as a major contester to the dark matter hypothesis for the explanation of these anomalies. However, in the years that followed, the theory attracted little attention and the dark matter hypothesis became the standard paradigm, eventually resulting in its incorporation in the standard cosmological model. Despite the diligent efforts by Milgrom and his collaborators to develop a coherent relativistic version of MOND  \citep{bekenstein2004,skordis2021}, Milgrom's theory, along with every other possible attempt of modifying general relativity and Newtonian gravity within dark matter physics was largely marginalized, and modified gravity as a possible explanation of dark matter phenomenology is still considered today by the vast majority of the scientific community as a non-viable alternative. 

The aim of this article is to perform a historical overview in order to understand the reasons why the dark matter hypothesis prevailed the hypothesis of modified gravity in the 1980s, despite the fact that some of the relevant observational anomalies could, in principle, be equally accounted for by postulating any one of the following two working hypotheses, which would eventually be incorporated into a more comprehensive scientific theory:

\begin{itemize}
	\item $H_1$: The existence of dark matter, a massive non-baryonic field which interacts with baryonic matter mainly via gravity
	\item $H_2$: The modification of standard Newtonian dynamics in the regime of low accelerations, and consequently of the theory of general relativity of which it is a non-relativistic limit
\end{itemize}

The hypothesis of dark matter would later be integrated in the $\Lambda$CDM model, the standard cosmological model about the universe that assumes the correctness of the theory of general relativity and requires the existence of cold dark matter and dark energy, each comprising about 27\% and 69\% of the total cosmic mass-energy budget respectively. The hypothesis of modified gravity was mainly embedded in various versions of MOND, a group of effective theories of gravity in which standard Newtonian dynamics cease to obtain below a critical acceleration constant. 

In recent literature, proponents of MOND theories from the physics community have often appealed to philosophical arguments to highlight the virtues of these theories by mainly focusing on their predictive success at the galactic scales, the (un)falsifiability of the dark matter hypothesis, and the non-detection of dark matter particles in colliders and direct searches \citep[e.g][]{merritt2017, merritt2020,sanders2019,milgrom2020,mcgaugh2021}. Several philosophers have also engaged with the discussion, leading to a recent surge of articles on the debate between dark matter and MOND. A characteristic example comes from the recent work of \citet{duerr2023} who carry out a rigorous theory evaluation of MOND and $\Lambda$CDM in terms of their respective ad-hocness, concluding that MONDian theories come out as severely ad hoc. A similar comparison is performed by \citet{martens2023} who focus on the explanatory structures of the two competing research programmes and the different explanatory ideals they seem to satisfy, such as simplicity and unification. In earlier work, \citet{massimi2018} tackled the debate between $\Lambda$CDM and MOND in terms of the different scales at which the two competing theories are empirically successful, highlighting, amongst other things, that the $\Lambda$CDM works best at large cosmological scales whereas MOND is more successful in the galactic regime, thus facing a `downscaling' and an `upscaling' problem respectively.\footnote{Further comparisons and discussions about the debate between $\Lambda$CDM and MOND can also be found in \citet{jacquart2021l} and \cite{debaerdemaecker2022}. The former is a discussion about whether the debate is best understood in terms of models or theories, and the latter concerns the impossibility of reconstructing a defence of MOND in terms of a meta-empirical assessment. \cite{martens2020a,martens2020b} take a different approach on the debate, by questioning the tenability of a strict conceptual distinction between space and matter.}

These works tackle the debate between $\Lambda$CDM and MOND as a problem of theory choice whose resolution is to be found in the various theoretical virtues of the two competing theories, such as their ad hocness, explanatory power, unificatory power, falsifiability etc. As such, they provide important insights on the theoretical virtues of MOND and the $\Lambda$CDM model (and the lack thereof), and have substantially contributed towards our understanding of the merits and shortcomings of both $\Lambda$CDM and MOND. However, they do not explicitly show why, in light of certain observations and theoretical developments by the 1980s, the scientific community almost in its entirety decided to pursue the hypothesis of dark matter as opposed to the hypothesis of modified gravity, \textit{regardless of the theory into which the latter would eventually be embedded}. The main goal of this article is to complement these works by answering precisely this question, i.e. by exploring and understanding the reasons behind the strong preference of the scientific community to invest its effort in the integration of dark matter into the Big Bang model and the development of models for dark matter candidates, as opposed to pursuing alternative theories of gravity for the explanation of dark matter related phenomena at the galactic and cosmological scale. 

The rationale behind undertaking this task is twofold. As we shall see, while physicists were gradually realising the presence of observational and theoretical anomalies from the 1960s until the early 1980s when the dark matter problem reached its pinnacle, the dilemma they were faced with was not one between two fully developed theories -- the $\Lambda$CDM and a modified gravity theory -- which would be resolved based on their theoretical virtues. Rather, what was on the table was a choice between two different working hypotheses which would be further pursued and eventually integrated into a more advanced and comprehensive theory.  Nevertheless, apart from MOND and its relativistic extensions which were pursued by a small number of physicists in the years to follow, no other theory of modified gravity that reproduces dark matter phenomenology without the requirement for dark matter has been seriously pursued since then.\footnote{Of course, any alternative theory would need to have a MOND-like behaviour as its limit. However, one can imagine the possibility of a yet-unconceived theory with a different framework that nonetheless replicates MOND's behaviour in low-acceleration regimes. Many thanks to an anonymous referee for pointing out this subtlety.}

It is therefore important to understand the reasons behind this decision of the scientific community, that is, to understand why physicists were much more inclined to develop dark matter models and experiments for the possible detection of dark matter particles, as opposed to developing further modified gravity theories which could in principle compete and even replace MOND as a plausible alternative to dark matter. And given that in the early 1980s neither dark matter nor MOND were fully developed, the decisive factors for the choice between the two working hypotheses are to be found not so much in the theoretical virtues of the two corresponding theories, but rather in the \textit{attractiveness} of the two competing hypotheses, or as it is often described in the literature, in their pursuitworthiness. Much of this article is devoted in identifying these factors, which we shall call the epistemic criteria for the pursuitworthiness of a hypothesis, although as will we shall see, the criteria also contain a strong pragmatic element. Some of the epistemic criteria that make a working hypothesis pursuitworthy sometimes overlap with the theoretical virtues of a more developed physical theory and hence the distinction between the theoretical virtues of a theory and the epistemic criteria for pursuitworthiness we are alluding to is not always clear. Nevertheless, the reasons for accepting a theory need not be the same as the reasons for pursuing a theory, and thus the analysis to be presented here is still illuminating in that it shows why dark matter superseded in the 1980s despite not being fully understood yet. Indeed, as we shall see, the reasons why dark matter was pursued were not exactly the same as the reasons why the dark matter was widely accepted based on precision measurements of the CMB and the observations of the Bullet cluster. We shall return to this issue in Section \ref{pursuit}.

The second motivation stems from the fact that an analogous situation where a choice between two working hypotheses is required arises in the context of dark energy. Similarly to the dark matter case, the postulation of dark energy as a new form of matter can, in principle, be dispensed with by adopting modified versions of general relativity such as $f(R)$ theories and scalar-tensor theories. However, unlike with MOND, the tolerance of the scientific community towards the pursuit of such classical modifications of general relativity is considerably higher, and physicists are much less reluctant to consider and pursue these alternatives. The natural question that arises is therefore, why physicists are significantly more receptive to the idea of modifying general relativity to dispense with the need for a dark energy field, as opposed to modifying general relativity to eliminate the requirement for dark matter.\footnote{It should be noted that in the dark energy case there are, in fact, more than two possible alternatives. The situation, however, still resembles the dark matter problem in that two of these possibilities concern the postulation of an exotic field and the modification of gravitational dynamics. These issues will be elucidated in Section \ref{darkenergy}. For a philosophical discussion on the pursuit of these alternatives in dark energy see \cite{wolf2024}.}

A plausible answer to this question will be given in terms of three epistemic criteria for the pursuitworthiness of a hypothesis: its problem-solving potential, its compatibility and feasibility of incorporation, and its independent testability. As will be shown, when considered jointly, these criteria make a compelling case for explaining the strong inclination of the scientific community to pursue the development of the dark matter hypothesis and its incorporation in the standard model for cosmology, as opposed to the development of possible modifications in standard gravitational dynamics. The main conclusion is that the preference of the scientific community towards the further pursuit of the dark matter hypothesis as opposed to modified gravity stems from the fact that the former hypothesis could solve more problems than its contester, could  be integrated with established scientific knowledge in a much easier and straightforward way and was independently testable. This however, is not the case with the two competing hypotheses in the dark energy case, which explains why classical modifications of general relativity are still considered as a viable alternative to the hypothesis of dark energy as an exotic field or a cosmological constant. These conclusions are in accordance with the sentiment by \cite{wolf2024} that no approach to the dark energy problem stands out as being superior in terms of pursuitworthiness.

In what follows, a brief historical overview of the most relevant observational and theoretical developments during the 1960s and 1970s is presented in Section \ref{state}, in order to fully appreciate the state of the art in the field of astrophysics by 1983 when Milgrom presented the first version of MOND. Section \ref{pursuit} will follow with a rational reconstruction of the context in which physicists decided to pursue the hypothesis of dark matter as opposed to the hypothesis of modified gravity based on the three aforementioned epistemic criteria. Finally, in Section \ref{darkenergy} a brief comparison with the problem of dark energy will be presented, showing why the pursuit of the modified gravity hypothesis in this case is somewhat more motivated and less challenging compared to the dark matter case.

	\section{State of the art in the 1980s}\label{state}
	
The establishment of dark matter theory is often portrayed as an inevitable result of accumulating evidence from high velocity dispersions in clusters and flat rotation curves in nearby galaxies. However, as also noted by \citet{deswart2017}, a better understanding of how the postulation of dark matter became a central component of the standard cosmological model requires a more holistic approach to the observational, theoretical and sociological developments that shaped the scientific landscape of the 1980s. As we shall see, this is when the postulation of dark matter as a non-baryonic field started gaining serious attention as a plausible hypothesis to be integrated in the existing scientific framework, which eventually led to the formulation of the $\Lambda$CDM model. Nevertheless, despite the growing consensus about the reality of dark matter by the early 1980s, Mordehai Milgrom decided to go against the prevailing trend of that era and pursue the hypothesis of modified gravity by developing the first version of MOND. To fully appreciate the boldness of Milgrom's attempt and the conditions under which the majority of the scientific community decided to go in the opposite direction, it is helpful to review the most important developments in astrophysics and cosmology in the preceding years.\footnote{The content of this section is largely drawn from the excellent historical analyses by \cite{sanders2010}, \cite{deswart2017}, \cite{bertone2018}, and \cite{peebles2020}, and the cited primary sources. While these articles certainly do not comprise the entirety of the scientific literature on dark matter-related research by the early 1980s, they nonetheless offer a comprehensive portrayal of the most significant scientific advancements during that period.}
	
\textit{Missing mass.} Famously, the first indication of the presence of non-luminous matter in the universe is traced back to the measurements of radial velocities in the Coma cluster by \cite{zwicky1933}. Using the virial theorem, Zwicky found that the galaxies in the outer regions of the cluster were moving much faster than expected given the amount of visible mass, from which he then concluded that some form of dark matter is present in these clusters and its mass is about 400 times larger than the mass of visible galaxies. As is well known, Zwicky's results were largely dismissed by the community and the discrepancy between the calculated mass and the observed mass was attributed to observational errors in accounting the mass and light of these galaxies. Several years later, the mass-to-light ratio in clusters and individual galaxies was re-examined independently by \citet{schwarzschild1954}, \cite{vandehulst1957} and \cite{oort1960} who all found clear evidence for an increasing mass-to-light ratio in the outer regions of the galaxy, implying a considerable discrepancy between luminous mass and gravitational mass.\footnote{As the name suggests, the mass-to-light ratio, $M/L$, indicates the proportion of the quotient between the total mass of a spatial volume (typically on the scales of a galaxy or a cluster) and its luminosity, measured in ergs per second per gram. For stellar systems, this ratio is typically expressed in solar units; e.g. a galaxy with $M/L=3$ has a mass-to-light-ratio that is 3 times larger than that of the sun. Galaxies typically have 2-5 times larger mass-to-light ratios compared to the sun, however, a mass-to-light ratio higher than 10, such as the ones derived in these studies, is difficult to achieve with normal stellar populations and requires the postulation of non-luminous matter.} Once again however, the results of these studies were received with caution, partly because in some cases, the authors themselves expressed their reservations about the validity of these results because of large uncertainties in the estimations of mass and light.\footnote{cf. \citet[p.281]{schwarzschild1954}:`This bewilderingly high value for the mass-luminosity ratio must be considered as very uncertain since the mass and particularly the luminosity of the Coma cluster are still poorly determined.'}

\textit{Extended rotation curves.} By the early 1970s, technological advancements in the field of radio astronomy eventually facilitated the -- until then impossible -- detection of fainter signals from continuous and spectral line emission from galaxies. These developments substantially improved the observations of the 21-cm line of hydrogen, an ideal probe of the distribution and motion of gas in spiral galaxies beyond their optical image, which were instrumental for the derivation of \textit{extended rotation curves}. These curves measure how the gas in galaxies rotates as a function of distance from the centre of the galaxy and hence provide a much more robust picture of the distribution of the total mass in the galaxies, compared to the rotation curves derived solely from their visible image.

One of the first measurements of extended rotation curves from 21-cm observations was published by \citet{rogstad1972} who found that the rotational velocities in five spiral galaxies rise sharply to a maximum value and then remain flat, confirming `the requirement for low-luminosity material in the outer regions of these galaxies' (p.320). Similar results were also obtained a few years later by \citet{roberts1975} who found that although hydrogen extends well beyond the optical image of galaxies, the rotational velocity of the gas is equal to the velocity of stars in the inner regions. Just as with the results of Schwarzchild and others, however, both works were largely dismissed as an effect of the poorly understood beam patterns of radio telescopes. In 1980, \citet{rubin1980} published precise spectroscopic observations of the rotation curves of 21 spiral galaxies using line emissions of hydrogen and nitrogen, in which all rotation curves once again appeared to be flat. These results were in agreement with earlier highly influential work by \citet{rubin1970} and \citet{freeman1970}, albeit with the difference that the rotation curves were now extended beyond the optical image of the galaxies, making a much more compelling case for the requirement of additional non visible mass.

\textit{Stability of spiral galaxies.} At about the same time as the first observations of extended rotation curves, the rapid development of computing power in the 1960s allowed the detailed study of the dynamics of galactic systems in simulations of Newtonian N-body systems. Amongst the pioneers of these early simulations of galactic systems were \cite{miller1968} and \cite{hohl1971} who studied the dynamics of spiral structures in rotationally supported disk galaxies, i.e. disks of particles resembling stars in equilibrium that are supported by their rotation about the centre of the galaxy, and thus the gravitational force pulling the stars towards the centre is balanced by the centrifugal force pushing them outwards. Contrary to the Newtonian expectation for a stable system, the simulations showed that the particle-stars were eventually switching from their initial circular orbits to highly elongated paths with large excursions in radius, suggesting that the system evolves from being rotationally supported to being pressure supported. The problem was that galaxies such as our very own Milky Way, which lived long enough to resemble this behaviour, did not look like pressure supported systems. Rather the rotation of stars around the centre of the galaxy appeared quite circular, indicating a rotationally supported system. 

Following these results, \cite{ostriker1973} showed that the presence of a massive halo in the outer regions of spiral galaxies provides the necessary stability. Interestingly, Ostriker and Peebles made no direct reference to the possibility of exotic (non-baryonic) dark matter making up this halo in their article. Rather, they speculated that the halo consists of `ordinary' low-luminosity objects such as white dwarfs and very low-mass stars, suggesting further observational searches ``to see if numerous very faint high-velocity stars exist in the solar neighbourhood'' (\textit{ibid.},p.480). This should not come as a surprise however, since the idea of non-baryonic dark matter was not particularly entrenched in the scientific community when Ostriker and Peebles published their results. The standard view at that time was that these anomalies are probably caused by the presence of a low-luminosity massive halo in galaxies, and such halos had already been proposed and reported in earlier works by \cite{oort1965} for instance, as well as in the aforementioned studies indicating that the mass-to-light ratio increases rapidly with distance from the centre of galaxies. The value of Ostriker and Peeble's achievement was in the realisation that this additional low-luminosity mass in terms of a `dark halo' also provides the required stability in the simulated galactic systems, already hinting towards the potential of the dark matter hypothesis to solve multiple problems at once.

\textit{Structure formation and CMB anisotropies.}  The discovery of Cosmic Microwave Background (CMB) radiation by \cite{penzias1979} in 1965 put an end to the steady-state model of cosmology and essentially established the Big Bang model. By the late 1960s, the consensus was that the structure and evolution of the Universe is described by the Big Bang model, according to which the universe began as an extremely hot and dense singularity about 13.8 billion years ago and has been expanding and cooling ever since. Following these developments, James Peebles was one of the first cosmologists to highlight the structure formation problem. In a series of papers, \citep{peebles1965,peebles1966,peebles1968} he showed that in order to produce the observed large-scale structure of the Universe, the original amplitude of the fluctuations in the photon-baryon fluid at the decoupling epoch must be relatively large, which would correspond to comparable fluctuations in the temperature of the CMB. However, such fluctuations were not observed in the temperature spectrum of the recently discovered CMB radiation, indicating the presence of a further anomaly in need of an explanation.\footnote{To elaborate, structure formation in a homogeneous universe can only happen from gravitational collapse if the density fluctuations are larger than the Jeans length, which is the distance travelled by a sound wave during a collapse timescale. However, before the decoupling of photons and baryons (at a redshift of $z\approx 1000$, or $\sim$ 300 000 years after the Big Bang) the speed of sound is comparable to the speed of light, which means that the Jeans scale is comparable to a causally connected region, i.e. an event horizon. Thus, during that time, smaller fluctuations which could potentially give rise to galaxies and clusters do not grow, but rather propagate as sound waves. Gravitational collapse can therefore only begin at the decoupling of photons and baryons, and with a relatively large amplitude ($\approx 10^4$) which should show up in the CMB temperature power spectrum. For a more detailed presentation of the structure formation problem see \citet[Ch.5]{peebles2020}.}

In parallel scientific developments at the time, physicists were starting to realise that the cosmic abundance of standard neutrinos is comparable to that of photons and that the former could, in fact, be massive. \cite{cowsik1973} and later \cite{szalay1976} were among the first to speculate that the cosmic abundance of neutrinos could provide the missing cosmological mass and explain Zwicky's missing gravitational mass in the observed clusters. These results were generalised a few years later by \cite{gunn1978} who pointed out that, not only standard model neutrinos, but \textit{any} heavy and stable non-interactive particle which is a cosmological relic is `an excellent candidate for the material in galactic halos and for the mass required to bind the great clusters of galaxies' (p.1015), speculating the existence of hypothetical undiscovered non-baryonic particles that could easily be linked with and motivated by parallel developments in particle physics predicting new particles.

Most importantly, the possible presence of non-baryonic relic particles pointed out by Gunn and his collaborators also had major implications for structure formation in the early universe, offering a potential solution to the structure formation problem. If in addition to photons and baryons, there exists a non-interactive fluid that dominates the matter budget of the Universe, then the sound speed in this fluid may be much lower than the speed of light, and therefore, early fluctuations in this fluid do not propagate but continue to grow. This means that these fluctuations can begin to collapse much earlier than the decoupling epoch, allowing the early formation of large scale structure, in accordance with the data from the CMB temperature spectrum. By the early 1980s, this potential solution to the structure formation problem was reported extensively by several people \citep{bond1983,bond1984,peebles1982,vittorio1984}.

\textit{The desire for a closed universe.} In addition to the structure formation problem, the possible presence of additional non-interactive matter in the Universe was also in sync with the view that the Universe is spatially closed, or at least flat, shared by many physicists at the time. Roughly speaking, according to the Friedmann equations there are three possible scenarios for the geometry of space depending on the value of the density parameter $\Omega$, the ratio between the actual density of matter and energy, $\rho$, and the critical density, $\rho_c$, required to balance the gravitational attraction of matter and the expansion of the Universe. A unit ratio corresponds to a flat geometry, a negative ratio corresponds to an open geometry, and a positive ratio (i.e. one in which the actual density is larger than the critical density) corresponds to a closed geometry in which the universe resembles a sphere. Even though this quantity was in principle measurable, no precise measurements were available by the 1980s to determine the shape of the universe, and the possibility of a closed universe was often presented in papers of the time as a strong preference `for essentially nonexperimental reasons' \citep[p.L1]{ostriker1974}.
	
What was well known however, was that, based on the observed abundances of deuterium and helium, the baryonic fraction of the cosmological critical density, $\Omega_b$, was significantly lower than the critical mass density required to have a closed universe \citep{gott1974}. This fact was directly connected to dark matter at about the same time in two influential papers by \cite{ostriker1974} and \cite{einasto1974}, who both highlighted that the masses of `ordinary galaxies' had been significantly underestimated and additional mass is required to reach the critical density. In particular, \citeauthor{einasto1974} pointed out that the total mass density of matter in galaxies is 20\% of the critical cosmological density, significantly less than the required mass for a closed universe. Similarly, \citeauthor{ostriker1974} famously concluded that the mass of galaxies `may have been underestimated by a factor of 10 or more' and that `if we increase the estimated mass of each galaxy by a factor well in excess of 10, we [...] conclude that observations may be consistent with a Universe which is “just closed” ($\Omega=1$)...' (\citeyear[p.L1]{ostriker1974}). The possible existence of a non-interactive fluid could therefore not only solve the structure formation problem, but also provide the required additional mass for a closed or flat universe---if this was indeed the right geometry.\footnote{Understanding the exact reasons for this strong preference for a closed universe in the 1970s and 1980s is an interesting project which is beyond the scope of this article, but nonetheless deserves to be studied in its own right. At first glance, the reasons for this preference seem to be---at least partially---related to the validation of Mach's principle (cf.\ \citet[p.30]{rindler1967}: `the choice of $k=1$ [denoting a positive curvature] might appear desirable. It implies closed space sections that would, in some sense, validate Mach's principle according to which the totality of matter in the universe and nothing else determines the local inertial frames.'). For a historical discussion of this issue see \cite{deSwart2020}.}

\textit{The renaissance and establishment 
of general relativity in cosmology.} In addition to these scientific developments in cosmology and astrophysics, one should also take into consideration the significant momentum that Einstein's general theory of relativity gained in the 1960s and 1970s, during the so-called `golden age' of general relativity \citep[pp.258-299]{thorne1995}. On the theoretical level, general relativity underwent a series of significant developments ranging from the formulations of the first singularity theorems by Penrose and Hawking \citep{penrose1965,hawking1970}, and Hawking's (\citeyear{hawking1976}) seminal work on black hole thermodynamics, to the earlier developments in new Hamiltonian formulations of the theory \citep{dirac1950,arnowitt1959,dewitt1967} and the important advancements in the understanding of gravitational waves \citep{pirani1957,bondi1962}.

These theoretical developments gradually led to the establishment of the general theory of relativity as the standard paradigm for the study of the universe in cosmology, leading to the so-called `cosmological turn'. Moreover, the link between theory and observation was especially reinforced after the first discovery of quasars in 1963 which provided rich empirical grounds for the -- until then heavily theoretical -- relativists to construct concrete physical models and eventually prove that the general theory of relativity was more than an abstract extension of Newtonian gravity in the strong field limit. By the end of the 1970s, general relativity was already empirically confirmed by a series of experiments confirming Einstein's equivalence principle (e.g.\ the Eötvös experiments, gravitational redshift experiments, the Hughes-Drever experiment, the Turner-Hill experiment, the Ives–Stilwell experiment, measurements of the constancy of fundamental constants), solar-system experiments measuring the values of post-Newtonian parameters (light deflection tests, time-delay effect, the perihelion of Mercury) as well as from the first evidence for gravitational radiation from the discovery of binary pulsars \citep{hulse1975,taylor1982}.\footnote{For a more comprehensive discussion on the experimental confirmation of general relativity see \cite{will1979,will2014}. For a historical overview of the renaissance of general relativity in general, see \cite{blum2020}.}

	\section{Pursuing a working hypothesis}\label{pursuit}
	
In light of these developments, the answer to the question why the hypothesis of dark matter was pursued with more force in the years to follow as opposed to the hypothesis of modified gravity already starts to become clearer. To some extent, the situation in the years that followed these scientific developments---i.e. from the early 1980s onwards---resembles what \citet[pp.109-114]{laudan1978} described as the `context of pursuit', an intermediary stage between the discovery and the justification of a scientific theory.

In Laudan's rational appraisal of the scientific practice, the discovery/proposal of a new theory or hypothesis is usually followed by the stage of pursuit, where scientists further investigate the theory or hypothesis with the aim of integrating it with a fully developed theory. Justification is the final stage by which the scientific community eventually rejects or accepts a theory as part of the established scientific knowledge.\footnote{cf. also \citet[p.252]{franklin1993}: `By discovery I mean the process by which a theory or hypothesis is generated and proposed. Pursuit is the further investigation of a theory or of an experimental result. Justification is the decision process by which the scientific community comes to accept or reject a theory or an experimental result as part of the corpus of scientific knowledge.'} One of Laudan's most important insights is that the context of pursuit must be distinguished from the context of acceptance since, often, the criteria by which scientists opt to pursue a hypothesis or a theory `might have nothing directly to do with the acceptability [...] of the theories in question' (\citeyear[p.110]{laudan1978}). Nonetheless, as \citet[fn.3]{franklin1993} notes, the pursuit of a hypothesis might indeed occur before, after, or even simultaneously to the justification of a theory and, in some cases, the same sort of evidence and reasons that make a hypothesis pursuitworthy can also justify it.\footnote{See also \cite{laudan1980}, and \cite{franklin1993}. For more recent discussions on pursuitworthiness see \cite{vsevselja2014}, \cite{dimarco2019}, \cite{lichtenstein2021}, and \cite{shaw2022}.}

In the context of dark matter-related research, the stage of discovery corresponds to what was described in the previous section where a number of anomalies and theoretical developments led to the proposal of, primarily, the hypothesis of dark matter, and, subsequently, the hypothesis of modified gravity by Milgrom in 1983. The main question that arises -- and which is the main focus of this article -- is what reasons led scientists to pursue the development of dark matter and its eventual integration to the $\Lambda$CDM model, as opposed to the modified gravity hypothesis. What follows is a plausible answer to this question based on the following epistemic criteria for the pursuitworthiness of a hypothesis: (a) its problem-solving potential, (b) its compatibility with established theories and the feasibility of incorporation, and (c) its independent testability. It should be stressed however, that these criteria are not suggested here as the best criteria that scientists \textit{should} use in order to pursue a hypothesis.\footnote{If anything, as \cite{laudan1978} and \cite{franklin1993} note, scientists can sometimes opt to pursue a hypothesis for whatever reason, even if they do not believe it is true, and it is not clear whether a definitive and optimal list of criteria can (and should) be compiled.} Rather, they have been identified based on a historical overview, as the ones that seem to have played the most decisive role in motivating scientists to pursue and further develop the dark matter hypothesis as opposed to modified gravity. They are thus presented here as part of a narrative to facilitate a better understanding of the underlying reasons why the pursuit of modified gravity theories for the explanation of dark matter phenomena has been largely neglected by the scientific community, as well as a basis for making a comparison with the dark energy case in the next section.

\textit{Problem-solving potential.} This is a widely discussed feature that often appears in discussions of unification and, as the name suggests, concerns the potential of a hypothesis to solve multiple problems at once.\footnote{\citet[pp.108-9]{laudan1978}, for instance, proposes problem-solving potential as the best criterion between competing theories and links it to the progressiveness of a theory. Pursuing a hypothesis that solves more problems amounts to making greater progress since less questions remain unanswered.} Arguably, one of the most decisive reasons for the prevalence of the dark matter hypothesis over the hypothesis of modified gravity was the greater problem-solving potential of the former. As shown in Section \ref{state}, the postulation of an additional type of non-interactive dark matter in the early 1980s had the potential to solve the problems of missing mass, the flat rotation curves, the problem of the stability of galaxies, and the structure formation problem. In addition, the possible presence of dark matter could also ease the philosophical worry that the Universe might not be spatially closed, providing, amongst other things, additional support to Mach's principle and other theoretical and empirical reasons to believe that the Universe was at least flat.

The motivating force of the problem-solving potential of the dark matter hypothesis is evident in the two articles by \cite{einasto1974} and \cite{ostriker1974} where the authors emphasise the fact that the postulation of dark matter not only solves the observational anomalies, but is also in sync with the requirement of additional mass to reach the critical mass density of the universe (e.g. \citet[p.L4]{ostriker1974}: `the great extent of rich clusters of galaxies [...] appear to indicate that ``$\Omega \approx 1$''. The arguments presented above indicate that the masses associated with ordinary spiral galaxies may make a cosmologically interesting contribution.') Moreover, it is even more evident in the articles linking the possible presence of dark matter with the solution of the structure formation problem, highlighting the fact that the dark matter hypothesis had the potential to solve problems that seemingly have a different origin, i.e. observational anomalies in the dynamics of galaxies and the formation of large structures in the Big Bang. For instance, when \cite{peebles1984} presents a dark matter model for the origin of galaxies, he explicitly mentions that `A strong additional motivation for this paper is the discovery that the model does have some interesting features that seem capable of accounting for major elements of the observational situation.' (p.470).

By contrast, the modified gravity hypothesis was, at least on the face of it, only able to provide a solution to problems directly related to the dynamics of galaxies, namely the missing mass problem in clusters, the flat rotation curves and the stability of spiral galaxies, although it was not entirely clear at the time how a single modification of general relativity in low accelerations could solve all three problems at once. In fact, Milgrom's first version of MOND was initially introduced as a possible solution to the first two problems, and it took him six more years to also articulate a possible solution to the problem of the stability of spiral galaxies, which should not come as a surprise given that MOND is a non-linear theory \citep{milgrom1989}. His remarks on the motivations for MOND in one of his first papers are rather illuminating: `...the success of the modified dynamics in explaining the dynamics in galaxies and galaxy systems [...] is the only justification for introducing it...' (\citeyear[p.369]{milgrom1983a}). Similarly, a previous, and less known, suggestion for a different modification of Newtonian gravity at long distances/low accelerations by \cite{finzi1963} was also introduced to address the problem of the missing mass in clusters, leaving the possibility of also solving `a number of other problems in different fields of astrophysics' (p.21) open. 

Nevertheless, compared to the dark-matter hypothesis, the hypothesis of modified gravity at the time had the disadvantage of not being able to say anything about the structure formation problem and the observed patterns of the CMB temperature, which is still considered by many, as one of the most major drawbacks of MOND \citep{dodelson2011}. In fact, later relativistic extensions of MOND, such as TeveS \citep{bekenstein2004} and RMOND \citep{skordis2021} have been largely developed with the aim of addressing this thorny issue, offering potential solutions to the large structure formation problem by accommodating the CMB power spectrum. Whether these proposed solutions are convincing remains a matter of debate within the scientific community; however, this issue is quite orthogonal to our discussion. What matters is that in the 1980s the problem of large structure formation could not be resolved by MOND or any other possible modification of gravity.

\textit{Compatibility with established theories and feasibility of incorporation.} Another important factor in pursuing a working hypothesis concerns how compatible it is with established scientific theories, and consequently with the data and observations upon which those theories have been tested. It should be stressed once again however, that this is not to say that only those hypotheses that are fully compatible with established theories should be further pursued. Rather, this criterion better aligns with the scientific practice when understood in negative terms, namely, as stating that those working hypotheses that are in direct conflict with theories that have been widely tested over time, are less likely to be pursued by the majority of the scientific community.

The hypothesis of dark matter as a field that mainly interacts via the gravitational force with ordinary luminous matter, had the major advantage of being fully compatible with the general theory of relativity, which as we have seen, by the early 1980s had been thoroughly tested experimentally and was considered by the community as the standard gravitational theory for cosmology. The postulation of dark matter did not require any modification to the theory since the gravitational field equations were already formulated in a way that accommodates the presence of \textit{any} form of energy-matter as a source of curvature in space time, including dark energy and dark matter. In the context of Friedmann cosmology, which was then considered as the most suitable model of general relativity for cosmology, dark matter is treated as another form of matter with its own energy density, contributing to the total mass-energy content of the universe.

The modified gravity hypothesis however, was by definition incompatible with general relativity since it presupposes a departure from standard gravitational dynamics in low accelerations. It is therefore likely that the incompatibility of MOND with the highly-esteemed and established theory of general relativity acted for many as an anti-motivational factor for its further pursuit. This is most evident from the fact that Milgrom's initial formulation of MOND was heavily criticised for violating fundamental principles of physics and general relativity such as the conservation of momentum and the equivalence principle. For instance, in a referee report on Milgrom's initial submission to \textit{Astronomy and Astrophysics Letters} the reviewer characteristically notes that `In this theory there are very considerable losses of accurately checked phenomena to achieve an interpretation of phenomena that are not well understood while maintaining that ``most of what there is can be seen'' and so dispensing with hidden matter.', and further continues by saying that this gain is obtained at the cost of various cherished principles such as equivalence and relativistic covariance.\footnote{As quoted in \citet[p.130]{sanders2015}.} As \citet[p.130]{sanders2015} notes, the referee's main point was that Milgrom is trying to save not-well established phenomena, at the cost of well-established physical principles, and until a more complete theory can be presented, Milgrom's work is not worthy of publication.

Nonetheless, compatibility in itself may not be a decisive factor for the pursuit of a hypothesis since it is always possible that a working hypothesis which, prima facie, seems incompatible with an established theory, is in fact compatible with a slightly modified version of the latter. In this case, the major focus is shifting from examining whether a hypothesis is compatible with established theories, to examining how feasible it is to incorporate the hypothesis into either an already established theory or into a new modified version of it. The feasibility of incorporation therefore concerns the practical dimension of integrating a working hypothesis in the established scientific knowledge and making it compatible with previous observations and experimental results outside the context in which it was initially proposed. In deciding whether to pursue a working hypothesis or not, scientists may therefore take into consideration whether it has realistic prospects of being incorporated into existing theories that have been thoroughly tested, or into a new theory which nonetheless will not be in tension with established scientific knowledge.

In this respect, the incorporation of the dark matter hypothesis, in a sense, came `for free', since no modifications whatsoever were required to integrate it into the $\Lambda$CDM model, which is essentially a combination of general relativity and the two postulates of cold dark matter and dark energy. What is more, the possible existence of dark matter particles was further motivated by the then recent developments in particle physics on early unified gauge theories in the 1970s, which were already predicting new particles that could be excellent candidates for dark matter.\footnote{cf. \citet[p.1015]{gunn1978}: `modern renormalizable unified gauge theories of the weak and electromagnetic interactions \citep {weinberg1974,veltman1973} have provided motivations for the existence of heavy leptons, both charged and neutral.'} The most characteristic example probably comes from supersymmetry, a promising extension of the standard model of particle physics which, amongst other things, had already provided a plausible dark matter candidate in terms of the lightest supersymmetric particle (also known as the neutralino). Dark matter was therefore not only easily integrated with general relativity, but also already incorporated into what appeared to be the most promising extension of the standard model of particle physics motivated by entirely independent theoretical considerations. In the years that followed, the proposed candidate models for dark matter particles grew dramatically, indicating the practical feasibility of integrating the dark matter hypothesis to the particle sector as well.\footnote{For a philosophical discussion on the proliferation of dark matter candidate models in particle physics and the challenges that arise therein see \cite{martens2022dark} and \cite{antoniou2023}.}

The integration of the modified gravity hypothesis with Newtonian dynamics and the theory of general relativity on the other hand was -- and still is -- a particularly challenging task. A modified theory of gravity addressing galactic dynamics should, at a minimum, appear as a natural limit of general relativity (or a modified version of it) at very low accelerations, while at the same time reproducing a vast array of well-tested gravitational phenomena at different scales. To accomplish this would seem to many a daunting task, especially given that the alternative route of dark matter was already on the table. For instance, in order to achieve a smooth transition between high and low acceleration regimes, MOND theories necessarily require an interpolation function which needs to be put in by hand and is therefore considered to be artificial and non-physical. Moreover, the fact that Milgrom's initial formulation of MOND was violating basic physical principles of physics such as the conservation of momentum and the equivalence principle simply shows how difficult his mission was. Milgrom managed to reproduce galactic phenomenology in great detail by modifying Newton's law, however, he only achieved this at the expense of some of the most basic and fundamental principles of physics.

One year after his initial publications, Milgrom collaborated with Jacob Bekenstein to develop AQUAL, a non-relativistic field theory of MOND, with the aim of addressing these theoretical problems \citep{bekenstein1984}. This attempt however, was also plagued by its own problems, the most important being the requirement of faster-than-light propagation of waves and the failure to reproduce and explain the phenomenology of gravitational lensing in its relativistic form. The requirement for faster-than-light propagation was also carried through to Bekenstein's later relativistic extension, TeVeS \citep{bekenstein2004}, which has been conclusively ruled out by the observation of gravitational waves in 2017 \citep{boran2018}.\footnote{For a critical philosophical discussion on the falsification of multimetric modified gravity theories see \cite{abelson2022}.} These issues are indicative of how much more challenging the task of integrating a modified gravity hypothesis into the rest of scientific knowledge was, compared to the hypothesis of dark matter.

\textit{Independent testability.} The third criterion which conceivably played an important role in pursuing dark matter over modified gravity is the prospect of independently testing a working hypothesis outside the domain from which it was initially proposed. Scientists may feel more inclined to pursue a hypothesis if it shows some promise of leading to novel predictions and can be experimentally tested based on evidence from different types of phenomena than the ones that led to its postulation in the first place. A possible explanation for this inclination can be found in the long-discussed issue of the greater confirmatory power of novel predictions, compared to the accommodation of already existing evidence. Hypotheses that only accommodate existing results and lack the prospective of being tested on new ground may be less pursued because of an underlying worry that scientists will not be able to convince the rest of the scientific community that they are true. By contrast, the prospect of making a novel prediction and the possibility of a triumphant confirmation of a working hypothesis by an independent experiment or observation can make a hypothesis much more attractive and further motivate its pursuit.\footnote{For further philosophical discussion and a Lakatosian take on this issue see the relevant work of \cite{merritt2021mond} and \cite{duerr2023}.}

The postulation of a non-interactive massive particle in the 1980s had clear empirical implications which could, at least in principle, be tested experimentally outside the domain of galactic dynamics. The fact that the postulated particles had to be massive and abundant, combined with the possibility that they interact weakly with baryonic mass was enough to envision several possible tests for the independent detection of such particles which eventually led to today's direct, indirect and collider searches for dark matter. Indeed, as early as in 1983, there were various discussions in the literature suggesting possible experimental tests for the detection of dark matter particles (cf. \citet[p.470]{peebles1984}: `There is no very strong evidence that globular clusters do have massive halos, but there are prospects for tests of halos at the wanted density' and \citet[p.1415]{sikivie1983}: `Experiments are proposed which address the question of the existence of the ``invisible'' axion for the whole allowed range of the axion decay constant.' and `axions may be the stuff the dark halos
of galaxies are made of.').

Even though such particles have not been detected yet, what matters is that the hypothesis of dark matter was clearly testable on independent grounds, as opposed to the rather vague hypothesis of modified gravity. In the absence of a fully developed alternative theory of gravity to support this hypothesis, the only justification for its introduction would come -- as it did in Milgrom's case -- from phenomenological considerations in galactic dynamics, without a clear way of formulating, at the time, any possible independent tests, let alone novel predictions.\footnote{Milgrom himself was in fact fully aware of this difficulty: `At the moment I cannot suggest a feasible laboratory experiment to test the ideas discussed above' (\citeyear[p.369]{milgrom1983a}).} Even today, the vast majority of possible tests for MOND-like theories suggested in the literature are exclusively based on data from galactic dynamics (i.e. within the domain in which MOND was originally developed) undermining its potential ability to make predictions in other domains and be subjected to independent tests \citep{iocco2015}.

In sum, we have argued based on historical considerations, that the facts that the dark matter hypothesis (a) was able to solve more problems than a possible modification of gravity, (b) was fully compatible with established knowledge in cosmology and particle physics and  could be easily integrated with general relativity and promising extensions of the standard model in particle physics, and (c) could potentially be tested by independent experiments, were jointly decisive in shifting the weight of research in the 1980s towards the further development of dark matter models. If considered individually, each one of these epistemic criteria may not seem to provide a sufficient reason to pursue a hypothesis, however, when jointly considered, the presented criteria  make a compelling case for understanding the decision of the majority of the scientific community to pursue dark matter back in the 1980s.

Given that the primary aim of this article is to explain the reasons why the dark matter hypothesis superseded the hypothesis of modified gravity in the 1980s, the focus of our analysis is, naturally, on the positive aspects that made the former a more attractive option than the latter. Nevertheless, it is also worth considering -- even briefly -- whether there were also reasons to reject the postulation of dark matter and pursue a modification of gravity. The most natural reason for doing so is the fact that dark matter, due to its exotic nature, was not part of the established knowledge about the fundamental constituents of matter at the time, namely the standard model of particle physics. This explains the initial scepticism of the community when Zwicky and others made the first claims about the possible existence of dark matter and the attribution of these results in measurement errors (without of course ignoring the fact that distances were indeed poorly determined at the time). The suggestion by \cite{ostriker1973} to search for `ordinary' low luminosity objects in the halos of galaxies in their seminal article on stability mentioned in Section \ref{state}, is also indicative of the fact that in the early 1970s the idea of dark matter as an exotic non-baryonic particle was not yet fully entrenched in the consciousness of the scientific community, which initially sought for a solution within the standard model of particle physics.

Moreover, Milgrom's main reason for rejecting the dark matter hypothesis was its ad hoc nature: `in order to explain the observations in the framework of this idea, one finds it necessary to make a large number of ad hoc assumptions concerning the nature of the hidden mass and its distribution in space.' \citep[p.365]{milgrom1983a}, even though his proposal was perhaps equally ad hoc.\footnote{cf. \cite[p.x]{merritt2017}: `Milgrom’s explanation for the rotation-curve anomaly is neither more nor less ad hoc than the dark matter postulate. Both are examples of what philosophers of science call `auxiliary hypotheses’: assumptions that are added to a theory in order to (in this case) reconcile it with falsifying data.'} One may therefore ask what Milgrom's main motivation was, and the answer is to be found on the ability of his hypothesis to solve some of the existing problems better than its competitor. That is, rather than solving all problems at once, Milgrom was interested in solving some of these problems (missing mass in clusters and flat rotation curves) in the best possible way without worrying if the hypothesis can also be applied elsewhere (i.e. to solve the large structure formation problem). The precision with which a hypothesis solves a particular problem, compared to another, can therefore also make it worth pursuing, as it did for Milgrom and Bekenstein in the 1980s. In a similar spirit, physicists currently working on MOND often highlight the tremendous success of the theory at the galactic scale while at the same time pointing towards several problems besetting the $\Lambda$CDM regarding galactic dynamics such as the cusp core problem, the missing satellite problem, the angular momentum catastrophe, and the problem of satellite planes.\footnote{See \cite{delpopolo2017} for a review of these problems.} 

Proponents of modified gravity theories have also voiced a number of additional concerns on the hypothesis of dark matter. The most common objection stems from the fact that dark matter has not been detected so far in experiments despite being much more abundant than baryonic matter. Moreover, even if dark matter indeed exists, galactic dynamics can also be fully determined by baryons, implying unexpected correlations between dark matter and baryonic mass.\footnote{See \cite[Sec. 4]{milgrom2020} for a review of these objections and references therein.} Most---if not all---of these arguments however, emerged only after the 1980s and thus fall beyond the scope of our analysis. For our purposes, it suffices to note that the non-detection of dark matter to date, as well as small-scale problems and other plausible concerns raised by advocates of MOND, were primarily developed only after the implications of the dark matter component in the $\Lambda$CDM were better understood. These arguments could not, therefore, have had a negative impact on the prevalence of dark matter in the 1980s, especially given the optimism of its future testability at the time. Possible explanations as to why these arguments have not shifted the focus towards a modified theory of gravity for explaining dark matter phenomenology today, can be found in the articles comparing $\Lambda$CDM with MOND in the present context, cited in the introduction of this paper.

As a final remark, let us also note that the eventual discovery of the missing primordial fluctuations first by the COBE satellite in 1991 and later by WMAP in the early 2000s, along with the observation of gravitational lensing phenomena in the Bullet cluster, marked, for many, the justification of the dark matter hypothesis, thus closing Laudan's circle of discovery, pursuit, and justification. The COBE and WMAP observations revealed for the first time the long sought-after fluctuations in the CMB temperature at the level necessary for the formation of structure given the presence of cold dark matter in the early Universe, and essentially cemented the hypothesis of dark matter as a necessary cosmic relic for the explanation of large-scale structure formation. A few years later, the detailed measurement of gravitational lensing effects in the Bullet cluster \citep{clowe2006} provided even stronger evidence for the existence of dark matter by showing that the separation of visible baryonic matter in the cluster can only be explained in terms of dark matter, and not by modifying gravitational dynamics. In the absence of a dark matter particle discovery in collider and direct searches to this day, the Bullet cluster provides, for many, the most conclusive evidence for the existence of dark matter.

Whether these two significant advancements sufficiently justify the acceptance of the dark matter hypothesis is largely a subjective matter which would require a separate argument, though the majority of physicists appear to support this view. Undoubtedly, the strongest form of justification will come from the detection of a dark matter particle in direct and collider experiments. However, if one accepts the view that the Bullet cluster phenomenology and the successful prediction of the CMB power spectrum justifies the hypothesis of dark matter, one clearly sees that the reasons for the acceptance of the dark matter hypothesis are clearly different than the reasons for its pursuit. In the latter case the motivations have a strong pragmatic element and are closely related to the potential prospects of the hypothesis, whereas in the former, the acceptance of the hypothesis is based on its novel predictions and the fact that a certain phenomenon necessarily requires the existence of some form of dark matter to be explained regardless of a possible modification of gravity. Whether these (or other) reasons constitute a rational basis for the acceptance of a hypothesis in the absence of direct empirical confirmation is an interesting question and deserves to be studied on its own merit in future work.

\section{The dark energy case}\label{darkenergy}
	
The dark energy problem in cosmology resembles to some extent the situation in dark matter, in that in the late 1990s the scientific community was once again faced with a dilemma between different working hypotheses which could, in principle, account for certain observational anomalies. Although the idea of dark energy has been present since Einstein's infamous addition of a cosmological constant to the field equations to accommodate the possibility of a closed and static universe, the modern version of the dark energy problem was mainly revived in 1998 when observational data from Supernovae Type Ia (SN Ia) showed strong evidence for an accelerating expansion of the Universe \citep{riess1998}. Similarly to the missing mass in clusters and the flat rotation curves, this observation highlighted the requirement for an explanation of this cosmic acceleration, which has since then been became known as `dark energy' and its exact nature still remains elusive.

The simplest solution to this problem was to identify the source of acceleration with a cosmological constant $\Lambda$, a free parameter in the theory of general relativity, whose energy density remains constant over time. In terms of compatibility and feasibility of incorporation, the integration of a physical constant to cosmological models is rather attractive since it can be very easily integrated into the Einstein field equations, and given that it is treated as a constant of nature, no further work is required to determine its physical properties.\footnote{In fact, some might argue that there is no external addition in the theory in this case: there was always a free parameter in the field equations in the form of an integration constant, and to set it to zero, rather including it, would require an argument.} Nevertheless, as also noted by \cite{smeenk2023}, the identification of the source of cosmological expansion with a physical constant is at the same time rather uninteresting and sterile, in that it does not generate any further empirical and theoretical consequences that can be independently tested. Moreover, this possible solution to the problem of dark energy is also beset by the so-called cosmological constant problem, which, roughly speaking, stems from the huge discrepancy of 121 orders of magnitude between the predicted theoretical value of the vacuum energy density in quantum field theory (with which the cosmological constant is identified) and the -- much smaller -- observed value of $\Lambda$.\footnote{Strictly speaking, the cosmological constant problem would persist even if $\Lambda$ were set to zero, since an explanation would still be required for why the vacuum energy of the quantum fields does not contribute to GR which would lead to an early collapse of the universe. For further philosophical discussions of the cosmological constant problem see \cite{schneider2020}, \cite{koberinski2021}, and \cite{koberinski2023}.} 

For these reasons, a large part of the scientific community finds the identification of dark energy with a cosmological constant unattractive and unmotivated, and has, since the late 1990s, also been pursuing two further hypotheses for the nature of dark energy: (i) the identification of dark energy with a modified form of matter and (ii) the identification of dark energy with modified gravity. The first hypothesis requires the postulation of an exotic form of matter with negative pressure that maintains a constant density and does not dilute as the Universe expands, counteracting the gravitational force and producing the accelerating expansion of the Universe. It is most often incorporated in models of quintessence \citep{tsujikawa2013}, k-essence \citep{armendariz2001} and dark energy as a perfect fluid \citep{kamenshchik2001}. The second hypothesis requires the modification of standard gravitational dynamics at the cosmological scale (i.e. the modification of general relativity) in a way that generates an accelerating cosmological expansion without the requirement of a cosmological constant or an exotic type of matter in the field equations. The most characteristic examples of modified gravity attempts to dispense with dark energy as matter or a constant, come from various scalar-tensor theories such as the Brans-Dicke theory \citep{brans1961}, $f(R)$ gravity \citep{sotiriou2010}, and various braneworld models \citep{dvali20004}. What is important however, is that the vast majority of these modifications  are required at the outset to satisfy local gravitational constraints, and as a result, do not have any direct empirical implications in gravitational phenomenology at low energies where dark matter related phenomena are observed.\footnote{To make the comparison between dark energy and dark matter clearer, the focus here will be on the choice between pursuing dark energy as modified matter and dark energy as modified gravity. However, for completeness, it should be mentioned that dark energy phenomenology can also be explained in terms of backreaction and void models of dark energy. The underlying common idea of these approaches is that the observed different expansion rates at different distances are not due to an accelerating universe, but are rather caused by a strong inhomogeneity. For a standard textbook exposition of the dark energy problem see \cite{amendola2010}.}

Just as with dark matter, the scientific community was therefore once again presented with two, in principle, promising working hypotheses to be integrated in a complete theory of the Universe (under the assumption that the cosmological constant $\Lambda$ is zero, or at least negligible). The major difference however, is that whereas in the dark matter case the vast majority of researchers pursued the hypothesis of dark matter, the situation with regards to dark energy is much more balanced in that roughly equal amounts of effort seem to be directed both towards the development of modified matter models and modified gravity models. The natural question that arises is therefore why, as opposed to the dark matter case, the hypothesis of modified gravity as a possible explanation for dark energy phenomenology is receiving considerably more attention compared to the hypothesis of modified gravity as a possible explanation for dark matter phenomenology. The answer to this question---or at least a partial answer---is found in the problem-solving potential of the two competing hypotheses, and most importantly in their compatibility and feasibility of incorporation with the established scientific knowledge.

To see why, the first thing to note is that the two competing hypotheses of dark energy as modified matter and dark energy as modified gravity are typically integrated with the general theory of relativity by modifying the right-hand side and the left-hand side of the Einstein field equations respectively. In the standard formulation of the Einstein equations:
\begin{equation*}
	G_{\mu\nu}=8\pi T_{\mu\nu}
\end{equation*}

the hypothesis of dark energy as modified matter in its various forms is typically incorporated by considering specific forms of the energy-momentum tensor $T_{\mu\nu}$ with negative pressure, whereas the hypothesis of dark energy as modified gravity is typically incorporated by modifying the Einstein tensor $G_{\mu\nu}$ in a way that reproduces the late-time accelerated expansion of the Universe without the requirement of an additional dark energy component. In this respect, and insofar as their possible quantum field implications are not taken into account, the two competing working hypotheses of dark energy as modified matter and modified gravity \textit{have no fundamental physical difference} from the point of view of general relativity, since one can always rephrase one hypothesis into the other by defining an appropriate conserved energy-momentum tensor that equals the Einstein tensor.\footnote{cf. \citet[p.5]{amendola2010}: `It is important to realize however that the two approaches, which we denote as modified matter and modified gravity, are not fundamentally different, at least if for a moment we do not consider their quantum field implications. From the viewpoint of classical General Relativity [...] one can always rephrase one into the other by defining a suitable	conserved energy-momentum tensor that equals the Einstein tensor.'} Unlike the dark matter case where (dark) matter and gravitational dynamics are treated as being fundamentally different within the Newtonian regime, the division of the two hypotheses in the dark energy case is mostly a practical way of categorising the two types of dark energy models into those that modify the energy-momentum tensor, and those that modify the Einstein tensor. Strictly speaking, within the theory of general relativity there is no way of distinguishing modified matter from modified gravity.\footnote{\cite{martens2020a} argue that the distinction between matter and gravitational dynamics is problematic also in the dark matter case, however, the problems they discuss appear once one enters the relativistic regime. Insofar as one starts from non-relativistic theories to account for dark matter related phenomena, as MOND does, the distinction between matter and dynamics is clear.}

The crucial difference compared to the dark matter case is, therefore, that both hypotheses for dark energy can be easily integrated with the general theory of relativity, whereas a hypothesis of modified gravity as an alternative to dark matter is much more difficult to incorporate.\footnote{This observation also aligns with the analysis in \cite{wolf2024} regarding the simplicity and conservatism of $f(R)$ gravity theories on the one hand, and the ease of model building for quintessence on the other.} The main reason is that in the former case one follows a top-down approach by starting with possible (and often simple) modifications of the field equations and working out their implications, whereas in the latter case, one follows a bottom-up approach by starting from low-acceleration phenomenology and then working out the possible modifications to Newtonian dynamics and subsequently general relativity, which, as we have seen, is a much more difficult task. This ease of integration is particularly evident in scalar-tensor theories and $f(R)$ theories, which are probably the simplest, and, as a result, the most intensely studied alternatives to general relativity. In scalar-tensor theories, the Ricci scalar $R$ couples to a scalar field $\phi$ with a coupling of the form $F(\phi)R$ in addition to the metric tensor field, and in $f(R)$ theories the 4-dimensional action from which the field equations are derived is given by some general function $F(R)$ of the Ricci scalar $R$ instead of the usual Einstein-Hilbert action in general relativity.

Another possible explanation for the balance in the pursuit of modified matter and modified gravity for the explanation of dark energy phenomenology can be found in the equal problem solving potential of the two hypotheses. Although the most significant problem the introduction of dark energy aims to solve is the accelerating expansion of the universe, this hypothesis also solves a number of additional problems of varying significance which were already known to cosmologists long before the observation of the acceleration of the Universe in 1998. Arguably, the two most important additional problems are the so-called `age problem' and the `critical density' problem. Leaving the details aside, the age problem amounts to the fact that if $\Lambda$ is taken to be zero, then the age of some astrophysical objects appears to be significantly older than the age of the universe. The critical density problem is that in order to achieve the critical density for a flat universe a large contribution from dark energy of $\Omega_{\Lambda}\approx 0.7$ is required in addition to the contribution from dark matter. Unlike the dark matter case where modifying gravity leaves some important problems unanswered, the fact that the two competing hypotheses of dark energy are essentially two faces of the same coin means that they are equally consistent with the relevant phenomenology, and hence their problem-solving potential is equivalent.\footnote{Here, it is also worth mentioning that recent constraints on the expansion rate of the Universe from Baryonic Acoustic Oscillations seem to render a particular class of quintessence models only marginally consistent with the data \citep{wolf2024scant}. The exact impact of these results on dark-energy research is yet unknown.}

Finally, two further reasons that differentiate the situation between dark matter and dark energy are worth mentioning. First, as opposed to the development of MOND which was motivated solely by dark matter phenomenology, classical modifications of general relativity which could potentially provide dark energy phenomenology such as the Brans-Dicke theory and $f(R)$ gravity had been already explored before the observation of SN Ia in 1998 for several reasons. For instance, the initial motivation for Brans and Dicke back in 1960s was the revival of Mach's principle, whereas $f(R)$ theories were mainly developed as toy theories to account for the non renormalizability of general relativity by complicating the action.\footnote{In more general terms, \cite{smeenk2023} mention three possible motivations for exploring alternatives to general relativity: (a) as possible ways to proceed to a successor theory, (b) to assess the rigidity of general relativity by showing that modifications lead to pathologies, and (c) to assess the robustness of empirical inferences from these modifications.} Hence, the requirement of dark energy from the accelerating Universe only reinforced the study of these alternatives by providing a further motivation.

Second, unlike the hypothesis of dark matter which, following the 1980s was strongly corroborated by the precision measurements of the CMB temperature by COBE and WMAP and the observation of the bullet cluster, the hypothesis of dark energy as modified matter still faces important pathologies which, so to speak, keep the hypothesis of dark energy as modified gravity alive. The most important of these pathologies are that the possible field mass of the dark energy particle is extremely small ($\approx10^{-33}eV$) compared to the typical masses appearing in the standard model of particle physics ($\approx10^6eV$), and that there is no convincing explanation for the absence of any coupling of this field to ordinary matter. This would also be the case for certain non-interacting candidate particles for dark matter (e.g. sterile neutrinos), however, the difference is that while such particles are only one of many viable candidates for dark matter---most of which are believed to be weakly interacting---there is no expectation for coupling in the case of dark energy.
	
	\section{Conclusions}
	
The presented analysis primarily aimed to explain why the pursuit of modified gravity as a plausible alternative to dark matter was largely neglected by the scientific community in the early 1980s, and subsequently to show why the situation in the dark energy case is different. An integrated history and philosophy of science approach to this question indicates that the pursuit of the dark matter hypothesis was mainly motivated by its greater problem-solving potential, its compatibility and feasibility of incorporation with established knowledge, and its prospects for independent testability. That is, postulating dark matter could potentially solve more problems than assuming different gravitational dynamics, while at the same time was aligned with the inclination of some astrophysicists towards a closed or flat geometry of the universe. Moreover, the working hypothesis of dark matter was fully compatible with the well established theory of general relativity, and somewhat already integrated into particle physics via early unified gauge theories, especially supersymmetry. Finally, the fact that such a non-baryonic field should be massive and probably weakly interactive also meant that it was in principle detectable via independent and realistic experiments.

In contrast, the modification of gravity to accommodate dark matter phenomenology was a considerably more challenging and thus, less attractive, task. In its initial conception, the hypothesis of modified gravity could only solve the problems related to galactic dynamics. Most importantly however, the hypothesis of modified gravity was much more difficult to incorporate into established knowledge and be made compatible with physical principles that had already been confirmed and widely accepted. These considerations, coupled with the fact that such a modification of gravity was not readily testable on independent grounds at the time, seem to have played a decisive role in shifting the focus of the scientific community towards the pursuit of dark matter from the 1980s onwards. Indeed, in the following years, research on dark matter expanded significantly, resulting in the development of numerous candidate models for dark matter particles and several experiments aimed at their detection. However, in the absence of a direct detection of a dark matter particle to date, the question of whether we have indeed transitioned to the context of justification following the discovery of the CMB and the Bullet cluster remains, for some, a contentious issue.

Finally, although the problems of dark matter and dark energy share certain similarities regarding their potential solutions, we have seen that the modification of gravity as an alternative to the introduction of a new form of matter to address the problem of dark energy, remains a viable option for the majority of the scientific community. This is primarily because dark energy as modified gravity and dark energy as modified matter both possess the same problem-solving potential, and the development and integration of the two competing hypotheses with general relativity is equally simple and straightforward. Moreover, whereas in the dark matter case one needs to start from low-acceleration phenomenology and work out the implications and required modifications to general relativity, modified gravity theories for dark energy are, in general, required from the outset to satisfy small-scale gravitational constraints and fundamental physical principles, which therefore makes their further development a much less challenging task compared to the development of a modified gravity theory that reproduces dark matter phenomenology. These facts largely explain why modified gravity as a possible explanation for dark energy is still pursued to an equal degree as a viable option, while modified gravity as a possible explanation for dark matter is not. 

\vspace{2cm}

\vspace{2cm}

\textbf{Acknowledgments:} I am greatly appreciative to Dennis Lehmkuhl, Noah Stemeroff, Erik Curiel, Michel Janssen and other members of the Lichtenberg group for history and philosophy of physics at the University of Bonn for stimulating discussions and constructive feedback on earlier versions of this work. Many thanks also to Karim Thebault, Ana Cretu and to the UPAC team at the University of Utrecht led by Niels Martens for their comments and	feedback on this work,  as well as to two anonymous referees. Research for this project was generously funded by the Humboldt foundation via a Humboldt postdoctoral fellowship at the University of Bonn.

	\newpage
	
	\bibliographystyle{chicago}
	\bibliography{DM_Vs_MG.bib}
	
\end{document}